%% file: isit2023_arxiv_version.tex
\documentclass[conference,a4paper]{IEEEtran}

\usepackage[utf8]{inputenc} 
\usepackage[T1]{fontenc}
\usepackage{url}              
\usepackage{cite}             

\usepackage[cmex10]{amsmath}  
\usepackage{mleftright}       
\mleftright                   

\usepackage{graphicx}         
\usepackage{booktabs}         

\usepackage{tcolorbox}

\usepackage{vmr-symbols-vecbold}
\usepackage{standard-macros}

\usepackage{glossaries}
\glsdisablehyper
\loadglsentries{glossary}

\newtheorem{theorem}{Theorem}

\newtheorem{corollary}{Corollary}
\newtheorem{definition}{Definition}

\newtheorem{remark}{Remark}

\newcommand{\data}{\text{\tiny data}}
\newcommand{\noise}{\text{\tiny noise}}
\newcommand{\ml}{\text{\tiny ML}}
\newcommand{\MSE}{{\rm MSE}}
\newcommand{\crb}{\text{\tiny CRB}}
\newcommand {\snr} {\mathtt{SNR}}

\def\comment#1{}

\newcommand{\stkout}[1]{
	\color{red}\ifmmode\text{\sout{\ensuremath{#1}}}\else\sout{#1}\fi\color{black}}

\newcommand{\addra}{}
\newcommand{\addraisit}{}
\newcommand{\delra}{\comment}

\IEEEoverridecommandlockouts 
\begin{document}

\title{A Bilateral Bound on the Mean-Square Error for Estimation in Model Mismatch} 



%
\author{%
  \IEEEauthorblockN{Amir Weiss$^{\star}$, Alejandro Lancho$^{\star}$, Yuheng Bu$^{\dagger}$, and Gregory W. Wornell$^{\star}$}
  \IEEEauthorblockA{$^{\star}$Massachusetts Institute of Technology \quad $^{\dagger}$University of Florida\\            
                    \quad\{amirwei, lancho, gww\}@mit.edu \quad\quad\quad buyuheng@ufl.edu
                    }
\thanks{
Alejandro Lancho has received funding from the European Union’s Horizon 2020 research and innovation programme under the Marie Sklodowska-Curie grant agreement No 101024432. This work is also supported by the National Science Foundation under Grant No CCF-2131115.}
}

\maketitle

\begin{abstract}
  A bilateral (i.e., upper and lower) bound on the mean-square error under a general model mismatch is developed. The bound, which is derived from the variational representation of the chi-square divergence, is applicable in the Bayesian and nonBayesian frameworks to biased and unbiased estimators. Unlike other classical MSE bounds that depend only on the model, our bound is also estimator-dependent. Thus, it is applicable as a tool for characterizing the MSE of a specific estimator. The proposed bounding technique has a variety of applications, one of which is a tool for proving the consistency of estimators for a class of models. Furthermore, it provides insight as to why certain estimators work well under general model mismatch conditions.
\end{abstract}
\vspace{-0.15cm}
\vspace{-0.2cm}
\section{Introduction}\label{sec:intro}
Classical bounds on the mean-square error (MSE) in parameter estimation traditionally assume that the statistical model, which describes the relation between the (random) observations and the parameter of interest, is fully known. Examples include the celebrated Cram\'er-Rao bound (CRB) for the nonBayesian framework \cite{kay1993fundamentals}, Van Trees (Bayesian CRB) \cite{van2004detection}, Barankin \cite{barankin1949locally}, Ziv-Zakai \cite{ziv1969some}, Abel \cite{abel1993bound} and Weiss-Weinstein \cite{weiss1985lower} bounds. For a more complete survey, see, e.g., \cite{todros2010generalA,todros2010generalB}.

While these classical bounds are key to understanding the fundamental limitations in optimal parameter estimation, they all refer to the case where the statistical model is \emph{exactly} faithful to the physics. A more realistic, and perhaps  contemporary approach acknowledges that the mathematical model does not precisely describe the true underlying physics, and attempts to account for this inherent, almost inevitable mismatch. Of course, this is particularly important from an engineering perspective, where approximations are often made, preferring a mismatch for the benefit of simplicity in implementation. 

In such cases, there is a need to understand the impact on performance. From this perspective, practical upper and lower bounds on the MSE under model mismatch are valuable, and as such have been recently receiving increasing attention \cite{fortunati2017performance}. More generally, aspects of this topic have a long history, for a variety of different forms of model mismatch and in a variety of different problems. For a selective list of representative examples, see, e.g., \cite{huber1967under,white1982maximum,vuong1986cramer,xu2004bound,verdu2010mismatched,fozunbal2010regret,fritsche2015cramer,fortunati2015lower,richmond2015parameter,diong2017generalized,pajovic2018misspecified,roemer2020misspecified,abed2021misspecified}.

While much of the focus has been on lower bounds (e.g., the misspecified CRB (MCRB) \cite{fortunati2016misspecified}), upper bounds are also important. For instance, it is often unclear when the MSE of an estimator derived under a different model from the true one will be bounded from above. \addraisit{While upper bounds on MSE have been considered for some (mismatch-free) special cases---including, e.g., \cite{seidman1968upper,timor1970upper,zakai1972lower,hawkes1976upper,ephraim1992lower,schniter2000bounds,belliardo2020achieving}---we are not aware of tools for upper bounding the MSE of a given estimator under general model mismatch conditions. }Such an upper bound would provide a  guarantee for the actual performance, even if with a gap from the exact performance, which is at any rate unknown since the true model is unknown. 

In this paper, we develop a useful bilateral (i.e., upper and lower) bound on MSE under a general model mismatch. The bound is applicable to both biased and unbiased estimators, and in both Bayesian and nonBayesian frameworks. It can be used for establishing the consistency of estimators for a class of models rather than for a single model. Furthermore, it has the potential to be useful in understanding why certain estimators are robust---i.e., work well for various models that deviate from the true underlying (and unknown) model.

\vspace{-0.2cm}
\section{Preliminaries and Background}\label{sec:preliminaries}
Let $\Theta\subseteq\reals^{K\times 1}$ be a parameter space, and $(\setX,\setF,\setP)$ be a complete probability space, where $\setX,\setF$ and $\setP=\{P_{\theta}:\rvectheta\in\Theta\}$ denote an observation (or sample) space, a $\sigma$-algebra on $\setX$, and a collection of probability distributions indexed by $\rvectheta\in\Theta$ over the common measurable space $(\setX,\setF)$, respectively. We assume that all the distributions $\{P_{\theta}\in\setP\}$ are absolutely continuous with respect to a measure $\mu$, which is assumed to be the Lebesgue measure unless stated otherwise. For brevity, we write $P_{\theta}$ simply as $P$, except where emphasis is required.

We now provide the necessary background for our results.
\vspace{-0.2cm}
\subsection{$f$-divergence and Variational Representation}\label{subsec:fdivergence}
An $f$-divergence is the following measure of discrepancy between  distributions, defined over a measurable space\cite{csiszar1967information}.
\begin{definition}[$f$-divergence]\label{definition1}
Let $P,Q\in\setP$ be two probability distributions on $\setX$, such that $P\ll Q$, namely, $P$ is absolutely continuous with respect to $Q$. Then, for any convex function $f:(0,\infty)\to\reals$ that: (i) is strictly convex at $1$; and (ii) $f(1)=0$, the $f$-divergence of $Q$ from $P$ is defined as
\begin{equation}\label{fdivergence}
     D_f(P||Q)\triangleq\Exop_Q\left[f\left(\frac{{\rm d}P}{{\rm d}Q}\right)\right],
\end{equation}
where $\frac{{\rm d}P}{{\rm d}Q}:\setX\to[0,\infty)$ denotes the Radon-Nikodym derivative of $P$
with respect to $Q$.
\end{definition}
An equivalent and useful form of \eqref{fdivergence} is by the variational representation of $f$-divergence (e.g., \cite{wu2017lecture}), which uses the notion of convex conjugation, defined as follows.
\begin{definition}[Convex conjugate]\label{definition2}
Let $f:(0,\infty)\to\reals$ be a convex function. The convex conjugate $f^*$ of $f$ is defined by
\begin{equation*}
     f^*(x)\triangleq \sup_{\lambda\in\reals}\, \big[\lambda x-f(\lambda)\big]\triangleq\,\sup_{\lambda\in\reals}\,\widetilde{f}(x,\lambda).
\end{equation*}
\end{definition}
Using the fact that a convex conjugate of a convex function is also convex, and that $(f^*)^*=f$, we obtain the variational representation of $f$-divergence \addra{\cite[Ch. 6.1.1]{wu2017lecture} }in terms of $f^*$,
\begin{equation}
     D_f(P||Q)=\sup_{g:\setX\to\reals} \Big[\Exop_P\left[g(\rndx)\right] - \Exop_Q\left[f^*\left(g(\rndx)\right)\right]\Big],\label{fdivvarrep}
\end{equation}
where $g$ is such that the expectations in \eqref{fdivvarrep} are finite.
\vspace{-0.1cm}
\subsection{Chi-square Divergence}\label{subsec:chisquaredivergence}
Specializing \eqref{fdivergence} with the function $f(x)=(x-1)^2$, we obtain the chi-square divergence (CSD, e.g, \cite{nielsen2013chi}),
\begin{equation*}
    \chi^2(P||Q)\triangleq\Exop_Q\left[\left(\frac{{\rm d}P}{{\rm d}Q}-1\right)^2\right]=\Exop_Q\left[\left(\frac{{\rm d}P}{{\rm d}Q}\right)^2\right]-1.
\end{equation*}
Using the fact that the convex conjugate of $f$ is given by
\begin{equation*}
    f^*(x)= \sup_{\lambda\in\reals}\, [\lambda x - (x-1)^2],
\end{equation*}
and after a relatively simple change of (the maximization) variable (see, e.g., \cite[Ch. 6.1]{wu2017lecture}), one obtains
\begin{equation}
    \chi^2(P||Q)=\sup_{g:\setX\to\reals} \frac{\left(\Exop_P\left[g(\rndx)\right]-\Exop_Q\left[g(\rndx)\right]\right)^2}{\Varop_Q\left(g(\rndx)\right)}\label{chisquaredivvarrep2},
\end{equation}
from which the Hammersley-Chapman-Robbins (HCR) \cite{hammersley1950estimating,chapman1951minimum} bound is readily derived \cite[Ch. 6.2]{wu2017lecture}. While the HCR pertains to a mismatch-free setting, we consider the fundamentally different model mismatch setting, as described next. 
\section{Estimation in Model Mismatch}\label{sec:mainresults}

Assume that the observations $\rndx_1,\ldots,\rndx_N$, denoted collectively as $\rvecx\triangleq\tp{[\rndx_1\,\cdots\,\rndx_N]}\in\setX$,\footnote{\delra{Conforming to}\addra{As in} traditional notation \delra{in}\addra{of} classical estimation, with a slight abuse of notation, we assume henceforth that $\rvecx$ (rather than $\rndx$) is an element in $\setX$.} are available for estimation of a vector of unknown parameters $\rvectheta\in\Theta$, and that $\rvecx$ and $\rvectheta$ are related via the model $P_{\data}\in\setP$. In a standard setting, $P_{\data}$ is assumed to be fully known, and given a proper criterion,\footnote{Be it for the frequentist or the Bayesian approach.} $\rvectheta$ can be estimated from $\rvecx$. Here, we consider the case where $P_{\data}$, the \emph{true} underlying relation between $\rvecx$ and $\rvectheta$, is (possibly partially) unknown. This setting is realistic in engineering problems where a physical model is unknown, or simply too complicated to describe analytically.

Since $P_{\data}$, the true relation between $\rvecx$ and $\rvectheta$, is in general unknown, the system designer chooses $Q_{\data}\in\setP$ to describe this relation. This choice is possibly based on some partial knowledge and/or simplifying approximations. Having chosen $Q_{\data}$, the system designer devises an estimator of $\rvectheta$ based on $\rvecx$, denoted by $\widehat{\rvectheta}(\rvecx)$, with an estimation error $\rvecvareps\triangleq\widehat{\rvectheta}(\rvecx)-\rvectheta$.

Note that while the estimator $\widehat{\rvectheta}(\rvecx)$ is designed based on $Q_{\data}$, its performance is affected by the true underlying model $P_{\data}$. Specifically, if we denote the resulting distribution of the \addra{squared }estimation error as $P\in\setP$, namely $\addra{\|}\rvecvareps\addra{\|_2^2}\sim P$, then the actual MSE of this estimator is given by
\begin{equation}\label{MSEunderP}
    \MSE_{P}\left(\widehat{\rvectheta}(\rvecx)\right)\triangleq \Exop_{P}\left[\|\rvecvareps\|_2^2\right]\in\reals_+.
\end{equation}
However, since $P_{\data}$ is unknown, the resulting $P$ is unknown, and therefore \eqref{MSEunderP} cannot be evaluated. Thus, the following natural (informal) questions arise: (i) Is it still possible to provide some performance guarantees in such (common) model mismatch situations? (ii) Is partial knowledge enough for some strong guarantees (e.g., consistency)? If so, (iii) how much knowledge about $P$ (or $P_{\data}$) is required to this end?

Our main result below shows that the answer to questions (i) and (ii) is yes. As for (iii), the more quantitative question, our results provide one possible answer that offers a trade-off between partial (reasonably available) knowledge and performance guarantees in terms of MSE (e.g., Corollary \ref{corollary2}).
\subsection{Main Results}\label{subsec:mainresults}
We now state our main result---a bilateral bound on the MSE in model mismatch.

\begin{theorem}\label{theorem1}
Let $\widehat{\rvectheta}(\rvecx)$ be an estimator of $\rvectheta$, and $\addra{\|}\rvecvareps\addra{\|_2^2}=\addra{\|}\widehat{\rvectheta}(\rvecx)-\rvectheta\addra{\|_2^2}$ its associated \addra{squared }estimation error\delra{ vector}, distributed according to $P$, stemming from the data distribution $P_{\data}$. Further, denote by $Q\in\setP$ the distribution of $\addra{\|}\rvecvareps\addra{\|_2^2}$, stemming from the chosen (possibly mismatched) data distribution $Q_{\data}$. Then, the true MSE \eqref{MSEunderP} is lower- and upper-bounded by
\tcbset{colframe=gray!90!blue,size=small,width=0.49\textwidth,halign=flush center,arc=2mm,outer arc=1mm}
\begin{IEEEeqnarray*}{l}
\MSE_{P}\left(\widehat{\rvectheta}(\rvecx)\right) \geq \MSE_{Q}\left(\widehat{\rvectheta}(\rvecx)\right) - \Delta\left(P,Q,\widehat{\rvectheta}(\rvecx)\right),
\IEEEyesnumber\label{bilateralbound}\IEEEyessubnumber\label{bilateralbound0} \\
\IEEEeqnarraymulticol{1}{r}{
\MSE_{P}\left(\widehat{\rvectheta}(\rvecx)\right) \leq \MSE_{Q}\left(\widehat{\rvectheta}(\rvecx)\right) + \Delta\left(P,Q,\widehat{\rvectheta} (\rvecx)\right)}, \IEEEyessubnumber\label{bilateralbound2}
\end{IEEEeqnarray*}
\noindent where $\Delta\left(P,Q,\widehat{\rvectheta}(\rvecx)\right)\triangleq\sqrt{\Varop_Q\left(\|\rvecvareps\|_2^2\right)\cdot\chi^2(P||Q)}\in\reals_+$.
\end{theorem}
\begin{IEEEproof}
It follows from \eqref{chisquaredivvarrep2} that
\begin{equation}\label{chisquaredivinequality}
    \chi^2\left(\delra{\widetilde{P}}\addra{P}||\delra{\widetilde{Q}}\addra{Q}\right)\geq \frac{\left(\Exop_{\delra{\widetilde{P}}\addra{P}}\left[g(\delra{\widetilde{\rvecx}}\addra{\rndx})\right]-\Exop_{\delra{\widetilde{Q}}\addra{Q}}\left[g(\delra{\widetilde{\rvecx}}\addra{\rndx})\right]\right)^2}{\Varop_{\delra{\widetilde{Q}}\addra{Q}}\left(g(\delra{\widetilde{\rvecx}}\addra{\rndx})\right)},
\end{equation}
for any $g(\cdot)$ for which the right-hand side in \eqref{chisquaredivinequality} is finite.

Now, choose $g(\delra{\cdot}\addra{t})=\delra{\|\cdot\|_2^2}\addra{t}$\delra{,} and $\delra{\widetilde{\rvecx}}\addra{\rndx}=\addra{\|}\rvecvareps\addra{\|_2^2}$\delra{ with}\addra{ such that} $P$ and $Q$\delra{, which} are the distributions of $\addra{\|}\rvecvareps\addra{\|_2^2}$\addra{,} induced by the data distributions $P_{\data}$ and $Q_{\data}$, respectively. For this particular choice, using the same definition \eqref{MSEunderP} for $Q$ as well, \eqref{chisquaredivinequality} specializes to    
\begin{equation*}
    \chi^2\left(P||Q\right)\geq \frac{\left(\MSE_{P}\left(\widehat{\rvectheta}(\rvecx)\right)-\MSE_{Q}\left(\widehat{\rvectheta}(\rvecx)\right)\right)^2}{\Varop_{Q}\left(\|\rvecvareps\|_2^2\right)}.
\end{equation*}
Multiplying both sides by (the positive) $\Varop_{Q}\left(\|\rvecvareps\|_2^2\right)$, taking the square root, writing the two resulting inequalities, and isolating $\MSE_P\left(\widehat{\rvectheta}(\rvecx)\right)$, readily gives \eqref{bilateralbound0}--\eqref{bilateralbound2}.
\end{IEEEproof}
\begin{remark}\label{remarkQterms}
In some cases, the terms that depend on $Q$ only (and not on $P$), namely $\MSE_{Q}\left(\widehat{\rvectheta}(\rvecx)\right)$ and $\Varop_{Q}\left(\|\rvecvareps\|_2^2\right)$, can be computed analytically, as they result from the chosen and \emph{known} data distribution $Q_{\data}$. Although $Q_{\data}$ may be different from the true data distribution $P_{\data}$, this can be very useful\addra{, e.g.,} for\delra{, e.g.,} proving consistency (see Corollary \ref{corollary2} below).
\end{remark}
\begin{remark}\label{bayesianornonBayesian}
A prior distribution may or may not be assigned to $\rvectheta$, and \eqref{bilateralbound} still holds\addraisit{ (see Section \ref{subsec:examplecalculation})}. Hence, the bound is applicable for both the Bayesian and nonBayesian frameworks.
\end{remark}
\begin{remark}\label{remarknomismatch}
In the absence of model mismatch, i.e., when the chosen data distribution accurately describes the data such that $Q_{\data}=P_{\data}$, the upper and lower bounds coincide, hence $\MSE_{Q}\left(\widehat{\rvectheta}(\rvecx)\right)=\MSE_{P}\left(\widehat{\rvectheta}(\rvecx)\right)$, as expected. While this property is trivial, it verifies that the bound \eqref{bilateralbound} is sensible. More generally, as long as the variance of the squared error is bounded under $Q$, the accuracy of the bound improves as the deviation (in the CSD sense) of $Q_{\data}$ from $P_{\data}$, and therefore of $Q$ from $P$, decreases. A desirable property, indeed.
\end{remark}
\begin{remark}\label{remarkoverlyoptimisticmodel}
The presumed, chosen model $Q_{\data}$ may be such that $\MSE_Q\left(\widehat{\rvectheta}(\rvecx)\right)\leq\MSE_P\left(\widehat{\rvectheta}(\rvecx)\right)$ (or even with strict inequality). However, \eqref{bilateralbound} shows that for such ``overly optimistic" choices of $Q$, a sufficiently high penalty must be incurred to the bound in the form of an increased $\Delta\left(P,Q,\widehat{\rvectheta}(\rvecx)\right)$ term. This \delra{will be}\addra{is} demonstrated \delra{in }\addra{via a simple example in }Section \delra{\ref{subsec:toyexample}}\addra{\ref{subsec:examplecalculation}}.
\end{remark}
\begin{remark}\label{remarkPunknown}
Recall that the MSE under $P$ cannot be computed, since $P$ is assumed to be partly or fully unknown. In this case, any nontrivial information regarding the actual MSE performance is valuable. As shown in Corollary \ref{corollary2} below, in some cases it is enough to (only) bound the $\chi^2\left(P||Q\right)$ term in order to attain nontrivial analytical performance guarantees.
\end{remark}

While the expectation and variance under $Q$ do not change when computed under $Q_{\data}$ (due to the ``law of the unconscious statistician'', e.g., \cite{ross2010first}), the CSD between the data distributions may be easier to compute or bound. This is owing to the fact that $Q$ and $P$ are transformed probability distributions, which are determined by the estimation rule $\widehat{\rvectheta}(\rvecx)$, and $Q_{\data}$ and $P_{\data}$. This motivates the following corollary.
\begin{corollary}\label{corollary1}
Consider the setting of Theorem \ref{theorem1}. Then, for (the true and presumed) distributions $P_{\data},Q_{\data}$ (respectively) of any function of the raw data, from which the estimator $\widehat{\rvectheta}(\rvecx)$ may be computed, we have
\begin{IEEEeqnarray*}{l}
\hspace{-0.15cm}\MSE_{P}\left(\widehat{\rvectheta}(\rvecx)\right) \geq \MSE_{Q}\left(\widehat{\rvectheta}(\rvecx)\right)\hspace{-0.05cm}-\hspace{-0.05cm}\Delta\left(P_{\data},Q_{\data},\widehat{\rvectheta}(\rvecx)\right),
\IEEEyesnumber\label{weakbilateralbound}\IEEEyessubnumber\label{lowerboundweak} \\
\IEEEeqnarraymulticol{1}{r}{
\hspace{-0.15cm}\MSE_{P}\left(\widehat{\rvectheta}(\rvecx)\right) \leq \MSE_{Q}\left(\widehat{\rvectheta}(\rvecx)\right)\hspace{-0.05cm}+\hspace{-0.05cm}\Delta\left(P_{\data},Q_{\data},\widehat{\rvectheta} (\rvecx)\right)}, \IEEEyessubnumber\label{upperboundweak}
\end{IEEEeqnarray*}
\end{corollary}
\begin{IEEEproof}
By the data processing inequality (e.g., \cite{csiszar1967information,csiszar1972class}),
\begin{IEEEeqnarray*}{l}
\chi^2\left(P||Q\right) \leq \chi^2\left(P_{\data}||Q_{\data}\right) \\
\qquad \Longrightarrow \Delta\left(P,Q,\widehat{\rvectheta}(\rvecx)\right)\leq \Delta\left(P_{\data},Q_{\data},\widehat{\rvectheta}(\rvecx)\right)\addra{.}\delra{,}
\end{IEEEeqnarray*}
\delra{hence}\addra{Since $\Varop_Q(\|\rvecvareps\|_2^2)=\Varop_{Q_{\data}}(\|\rvecvareps\|_2^2)$,} \eqref{lowerboundweak} and \eqref{upperboundweak} bound from below and above \eqref{bilateralbound0} and \eqref{bilateralbound2}, respectively.
\end{IEEEproof}

While \eqref{weakbilateralbound} is weaker than \eqref{bilateralbound}, as explained in Remark \ref{remarkPunknown}, it may be more convenient to work with; approximating the estimation error distribution is less trivial than approximating the distribution of the data, since in general we have access to data. Further, \eqref{weakbilateralbound} may already provide a satisfactory bound.

\addraisit{\subsection{The Gaussian Signal Model}\label{subsec:gaussianmodel}}
A particular case of high interest is when $Q_{\data}$ is chosen as the Gaussian distribution. Indeed, in many applications, such as communication, localization and image denoising, the Gaussian signal model is used in order to derive different estimators for various purposes, even though the actual measured signals clearly do not follow a Gaussian distribution. However, despite the model mismatch, many of these methods work well on real data. This suggests that there exists a more fundamental justification \delra{to}\addra{for} this fact than simply a good empirical fit.

Theorem \ref{theorem1} provides such a justification, and in particular, an accurate analytical description to the class of models for which the (mismatched) Gaussian model still yields ``good'' (e.g., consistent) estimators. Consider, for example, the general, ubiquitous signal model (e.g., as in \cite{yeredor2018high})
\begin{equation}\label{signal_in_noise}
    \rvecx_n = \rvech(\rvectheta) + \rvecv_n\in \reals^{M\times 1}, \quad n\in\{1,\ldots,N\},
\end{equation}
where $\rvech:\reals^{K\times 1}\to\reals^{M\times 1}$ is a known \delra{deterministic}\addra{(possibly random)} function, $\rvectheta$ is deterministic and unknown, and $\{\rvecv_n\}$ are zero-mean additive noise vectors with an unknown distribution\delra{ $P_{\noise}$}. If we nonetheless choose to assume that $\delra{\rvecv}\addra{\rvecx}_n\overset{\text{iid}}{\sim} Q_{\delra{\noise}\addra{\data}}=\setN(\delra{\rveczero}\addra{\rvecmu(\rvectheta)},\matLambda\addra{(\rvectheta)})$,\footnote{The symbol $\overset{\text{iid}}{\sim}$ stands for independent, identically distributed (iid).} we may derive $\widehat{\rvectheta}_{\ml}^{\setN}$, the maximum-likelihood estimator (MLE) of $\rvectheta$ for this chosen model.{\comment{ \footnote{which is also the least-squares (LS) estimator of $\rvectheta$ in this specific case.}}} However, the performance of $\widehat{\rvectheta}_{\ml}^{\setN}$, which is not necessarily the MLE for data $P_{\data}$ when $P_{\delra{\noise}\addra{\data}}\neq Q_{\delra{\noise}\addra{\data}}$, is generally no longer necessarily appealing. We emphasize that $\{\delra{\rvecv}\addra{\rvecx}_n\}$ are generally \emph{not} even iid under $P_{\delra{\noise}\addra{\data}}$. The following corollary describes the \delra{noise}\addra{data} distributions for which $\widehat{\rvectheta}_{\ml}^{\setN}$ is consistent despite model mismatch.
\begin{corollary}\label{corollary2}
Consider the signal model \eqref{signal_in_noise} and assume that $\rvech$ is such that $\widehat{\rvectheta}_{\ml}^{\setN}$ is consistent under $\addra{\rvecx_n\overset{\text{iid}}{\sim}}Q_{{\delra{\noise}\addra{\data}}}=\setN(\delra{\rveczero}\addra{\rvecmu(\rvectheta)},\matLambda\addra{(\rvectheta)})$, i.e., \delra{$\rvecx_n\overset{\text{\emph{iid}}}{\sim} Q_{\data}=\setN(\rvech(\rvectheta),\matLambda)$ and }$\widehat{\rvectheta}_{\ml}^{\setN}\xrightarrow[\quad]{p}\rvectheta$ as $N\to\infty$\addraisit{, where $\xrightarrow[\quad]{p}$ denotes convergence in probability}. Define $\smash{\Bar{\rvecx}\triangleq\frac{1}{N}\sum_{n=1}^N\rvecx_n}$, and denote $\Bar{\rvecx}\delra{-\rvech(\rvectheta) }\sim \Bar{P}_{{\delra{\noise}\addra{\data}}}$ and $\Bar{Q}_{{\delra{\noise}\addra{\data}}}\triangleq\setN(\delra{\rveczero}\addra{\rvecmu(\rvectheta)},N^{-1}\cdot\matLambda\addra{(\rvectheta)})$. If $\Bar{P}_{{\delra{\noise}\addra{\data}}}, \Bar{Q}_{{\delra{\noise}\addra{\data}}}$ are ``not too far'', i.e.,\footnote{\addraisit{The ``little-o'' notation $a_n = o(b_n)$ means that $\lim_{n\to\infty}a_n\cdot b^{-1}_n=0$.}}
\begin{equation}\label{conditionondata}
    \chi^2\left(\Bar{P}_{{\delra{\noise}\addra{\data}}}||\Bar{Q}_{{\delra{\noise}\addra{\data}}}\right)=o(N^2),
\end{equation}
then the estimator $\widehat{\rvectheta}_{{\ml}}^{\setN}$ is MSE-consistent under $P_{{\data}}$, namely,
\begin{equation}\label{mseconsistency}
    \lim_{N\to\infty} \MSE_P\left(\widehat{\rvectheta}_{{\ml}}^{\setN}\right)=0.
\end{equation}
\end{corollary}
\begin{IEEEproof}
For the model \eqref{signal_in_noise}, under $\rvecx_n\overset{\text{iid}}{\sim} Q_{\data}$, a sufficient statistic is $\Bar{\rvecx}\sim\setN(\delra{\rvech}\addra{\rvecmu}(\rvectheta),N^{-1}\cdot\matLambda\addra{(\rvectheta)})\triangleq\Bar{Q}_{\data}$. Therefore, $\Bar{\rvecx}$ can be treated as the observed data, with $\Bar{Q}_{\data}$ and $\Bar{P}_{\data}$ as its hypothesized (possibly mismatched) and true distributions, respectively. From Corollary \ref{corollary1}, we have
\begin{equation*}
\MSE_P\left(\widehat{\rvectheta}_{\ml}^{\setN}\right)\leq \MSE_Q\left(\widehat{\rvectheta}_{\ml}^{\setN}\right) + \Delta\left(\Bar{P}_{\data},\Bar{Q}_{\data},\widehat{\rvectheta}_{\ml}^{\setN}\right).
\end{equation*}
While we use $\widehat{\rvectheta}_{\ml}^{\setN}$ for brevity, we emphasize that in this case we have $\widehat{\rvectheta}_{\ml}^{\setN}(\Bar{\rvecx})$, since $\Bar{\rvecx}$ is a sufficient statistic. Now, since $\widehat{\rvectheta}_{\ml}^{\setN}$ is asymptotically efficient under $Q_{\data}$,
\begin{equation}\label{asymptoticdistributionMLE}
    \rvecvareps=\widehat{\rvectheta}_{\ml}^{\setN}-\rvectheta\xrightarrow[\quad]{d}\setN\left(\rveczero,N^{-1}\matI^{-1}(\rvectheta)\right),
\end{equation}
where \addraisit{$\smash{\xrightarrow[\quad]{d}}$ denotes convergence in distribution, $\rveczero$ is the all-zeros vector (with proper dimensions), and} $\matI(\rvectheta)\addraisit{\in\reals^{K\times K}}$ is the Fisher information matrix\addraisit{ (see, e.g., \cite[Ch. 11.10]{cover1999elements})} of a single observation $\rvecx_n$, hence
\begin{equation*}
    \MSE_Q\left(\widehat{\rvectheta}_{\ml}^{\setN}\right)=\frac{1}{N}\sum_{k=1}^K\underbrace{\left[\matI^{-1}(\rvectheta)\right]_{kk}}_{\triangleq\sigma_{\crb,k}^2}\xrightarrow[]{N\to\infty}0.
\end{equation*}

It remains to show that $\Delta\left(\Bar{P}_{\data},\Bar{Q}_{\data},\widehat{\rvectheta}_{\ml}^{\setN}\right)\xrightarrow[]{N\to\infty}0$\delra{. For this, observe first that, since $\Exop_{\Bar{P}_{\data}}[\Bar{\rvecx}]=\Exop_{\Bar{Q}_{\data}}[\Bar{\rvecx}]=\rvech(\rvectheta)$, then}\addra{, where}
\delra{whence}
\begin{equation*}
    \Delta\left(\Bar{P}_{\data},\Bar{Q}_{\data},\widehat{\rvectheta}_{\ml}^{\setN}\right)=\sqrt{\Varop_{\Bar{Q}_{\data}}\left(\|\rvecvareps\|_2^2\right)\cdot\chi^2\left(\Bar{P}_{\delra{\noise}\addra{\data}}||\Bar{Q}_{\delra{\noise}\addra{\data}}\right)}.
\end{equation*}
Focusing on the variance of the squared (norm of the) error,
\begin{align}\label{varianceofsquarederror}
    \Varop_{\Bar{Q}_{\data}}\left(\|\rvecvareps\|_2^2\right)&=\sum_{k=1}^K \Varop_{\Bar{Q}_{\data}}\left(\varepsilon_k^2\right)+ \sum_{\substack{k,\ell=1 \\ k\neq\ell}}^K {\rm COV}_{\Bar{Q}_{\data}}\left(\varepsilon_k^2,\varepsilon_{\ell}^2\right)\delra{.}\addra{,}
\end{align}
where ${\rm COV}_{\Bar{Q}_{\data}}(\rnda,\rndb)$ denotes the covariance of $\rnda$ and $\rndb$\addra{, evaluated under $\Bar{Q}_{\data}$}. Due to \eqref{asymptoticdistributionMLE}, we have, asymptotically,
\begin{equation}\label{varianceofnormaldistribution}
\Varop_{\Bar{Q}_{\data}}\left(\varepsilon_k^2\right)=\underbrace{\Exop\left[\varepsilon_k^4\right]}_{3\frac{\sigma_{\crb,k}^4}{N^2}}-\underbrace{\Exop^2\left[\varepsilon_k^2\right]}_{\left(\frac{\sigma_{\crb,k}^2}{N}\right)^2}=\frac{2\sigma_{\crb,k}^4}{N^2},
\end{equation}
where we have used Isserlis' theorem \cite{isserlis1918formula} to compute $\Exop\left[\varepsilon_k^4\right]$. As for the covariance terms in \eqref{varianceofsquarederror}, applying the Cauchy-Schwarz inequality and \eqref{varianceofnormaldistribution}, we obtain
\begin{align*}
{\rm COV}_{\Bar{Q}_{\data}}\left(\varepsilon_k^2,\varepsilon_{\ell}^2\right)&\leq \sqrt{\Varop_{\Bar{Q}_{\data}}\left(\varepsilon_k^2\right)\Varop_{\Bar{Q}_{\data}}\left(\varepsilon_{\ell}^2\right)}\\
&=\frac{2\sigma_{\crb,k}^2\sigma_{\crb,\ell}^2}{N^2}.
\end{align*}
Therefore, an upper bound on \eqref{varianceofsquarederror} is
\begin{equation}\label{upperboundonvariance}
    \Varop_{\Bar{Q}_{\data}}\left(\|\rvecvareps\|_2^2\right)\leq\frac{2}{N^2}\underbrace{\|\matSigma\|_{{\rm F}}^2}_{\text{independent of $N$}},
\end{equation}
where $\|\cdot\|_{{\rm F}}$ denotes the Frobenius norm and the entries of the auxiliary matrix $\matSigma\in\reals^{K\times K}$ are defined as
\begin{equation}\label{fourthordermoment}
 [\matSigma]_{k\ell}\triangleq\begin{cases}
 \sigma_{\crb,k}^2, & k=\ell,\\
 \sigma_{\crb,k}\sigma_{\crb,\ell}, & k\neq\ell.
 \end{cases}
\end{equation}
Using the bound \eqref{upperboundonvariance} and assumption \eqref{conditionondata}, we conclude that
\begin{equation*}
    \Delta\left(\Bar{P}_{\data},\Bar{Q}_{\data},\widehat{\rvectheta}_{\ml}^{\setN}\right)\leq\sqrt{\frac{2\|\matSigma\|_F^2}{N^2}\cdot\chi^2\left(\Bar{P}_{\delra{\noise}\addra{\data}}||\Bar{Q}_{\delra{\noise}\addra{\data}}\right)}\xrightarrow[]{N\to\infty}0,
\end{equation*}
and \eqref{mseconsistency} follows.
\end{IEEEproof}
\begin{remark}
For simplicity, in the proof of Corollary \ref{corollary2} we assume that the ``standard'' regularity conditions hold (e.g., \cite{lehmann2006theory}), such that $\widehat{\rvectheta}_{\ml}^{\setN}$ under $Q_{\data}$ is asymptotically efficient, and the Fisher information matrix exists. However, consistency (rather than asymptotic efficiency) of $\widehat{\rvectheta}_{\ml}^{\setN}$ can suffice for \eqref{mseconsistency}.
\end{remark}
It is interesting to note that, similarly to the requirement \eqref{conditionondata}, the Barankin bound also requires a (finiteness) condition on the CSD \cite[Eq.~6]{glave1972new} (though different in nature).

The importance of Corollary \ref{corollary2} is that it provides an analytical characterization, in the form of a sufficient condition, for the success of the Gaussian quasi-ML approach \cite{white1982maximum} in terms of MSE-consistency. Put simply, the condition \eqref{conditionondata} means that as long as the \delra{noise}\addra{data} distribution is not ``too far'' from the Gaussian distribution, using the Gaussian MLE\delra{, or equivalently the LS estimator,} is a reasonable approach for estimation, at least asymptotically.

Note further that the proof can be generalized for a non-Gaussian $Q_{\delra{\noise}\addra{\data}}$, as long as a similar condition as \eqref{conditionondata} holds. Moreover, theoretically, it is possible to take the infimum of \eqref{upperboundweak} over the parameters of the chosen $Q_{\delra{\noise}\addra{\data}}$---for example, in the Gaussian case, over the mean vector $\rvecmu(\rvectheta)$ and the positive-definite covariance matrix $\matLambda(\rvectheta)$---to get the tightest upper bound of this type.
\addra{\vspace{-0.1cm}
\subsection{Example Calculation}\label{subsec:examplecalculation}
The following example, which permits an analytical calculation of the bound, demonstrates aspects of the bound behavior. In particular, we consider the optimal multi-sensor receiver with angular mismatch. More specifically, consider a Bayesian version of \eqref{signal_in_noise}, with $\rvech(\rvectheta)=\rveca(\varphi)\rnds$, where $\theta=\rnds\sim\mathcal{N}(0,1)$ is the estimand (i.e., $K=1$ and $N=1$) and $\rveca(\varphi)\in\reals^{M\times 1}$ is a (unit-norm) steering vector as a function of the direction-of-arrival (DOA) $\varphi\in[0^{\circ},180^{\circ})$, and $\rvecv\sim\mathcal{N}(\rveczero,\frac{1}{\snr}\matI_M)$.\footnote{\addra{This example in fact deals with the linear minimum MSE (LMMSE) receiver \cite{tse2005fundamentals} of a single-input multiple-output (SIMO) communication system.}} In this case, the minimum MSE (MMSE) estimator of $\rnds$ is given by $\widehat{\rnds}(\varphi,\snr)\triangleq\frac{\snr}{1+\snr}\tp{\rveca(\varphi)}\rvecx$, when $\varphi$ is known. In practice, however, $\varphi$ is known only up to some accuracy level. Consequently, if some $\widetilde{\varphi}$ is used in place of $\varphi$, then\footnote{Here, $P$ and $Q$ are the error (rather than the squared error) distributions. Thus, by virtue of Corollary \ref{corollary1}, we consider a slight variant of Theorem \ref{theorem1}.}
\begin{align*}
    &\varepsilon=\widehat{\rnds}(\widetilde{\varphi},\snr)-\rnds\sim\hspace{-0.025cm}\begin{cases}
    P\hspace{-0.05cm}=\hspace{-0.05cm}\setN\left(0,\sigma^2\right), & \hspace{-0.3cm}\text{for }P_{\data}\leftrightarrow\widetilde{\varphi}\neq\varphi,\\
    Q\hspace{-0.05cm}=\hspace{-0.05cm}\setN\left(0,\frac{1}{1+\snr}\right), & \hspace{-0.3cm}\text{for }Q_{\data}\leftrightarrow\widetilde{\varphi}=\varphi,\\
    \end{cases} \\
    &\underbrace{\sigma^2}_{\MSE_P\left(\widehat{\rnds}\right)}\hspace{-0.075cm}\triangleq\hspace{-0.075cm}\underbrace{\tfrac{1}{1+\snr}}_{\MSE_Q\left(\widehat{\rnds}\right)}\cdot\underbrace{\left[3-2\tp{\rveca(\varphi)}\rveca(\widetilde{\varphi})+\tfrac{\left(\snr(1-\tp{\rveca(\varphi)}\rveca(\widetilde{\varphi}))\right)^2}{1+\snr}\right]}_{\triangleq1/\gamma^2(\snr,\varphi,\widetilde{\varphi})\geq\;1\;\text{ (``inflation factor'')}}\hspace{-0.05cm}.
\end{align*}
}
\begin{figure}[t!]
\centering
\includegraphics[width=0.95\columnwidth]{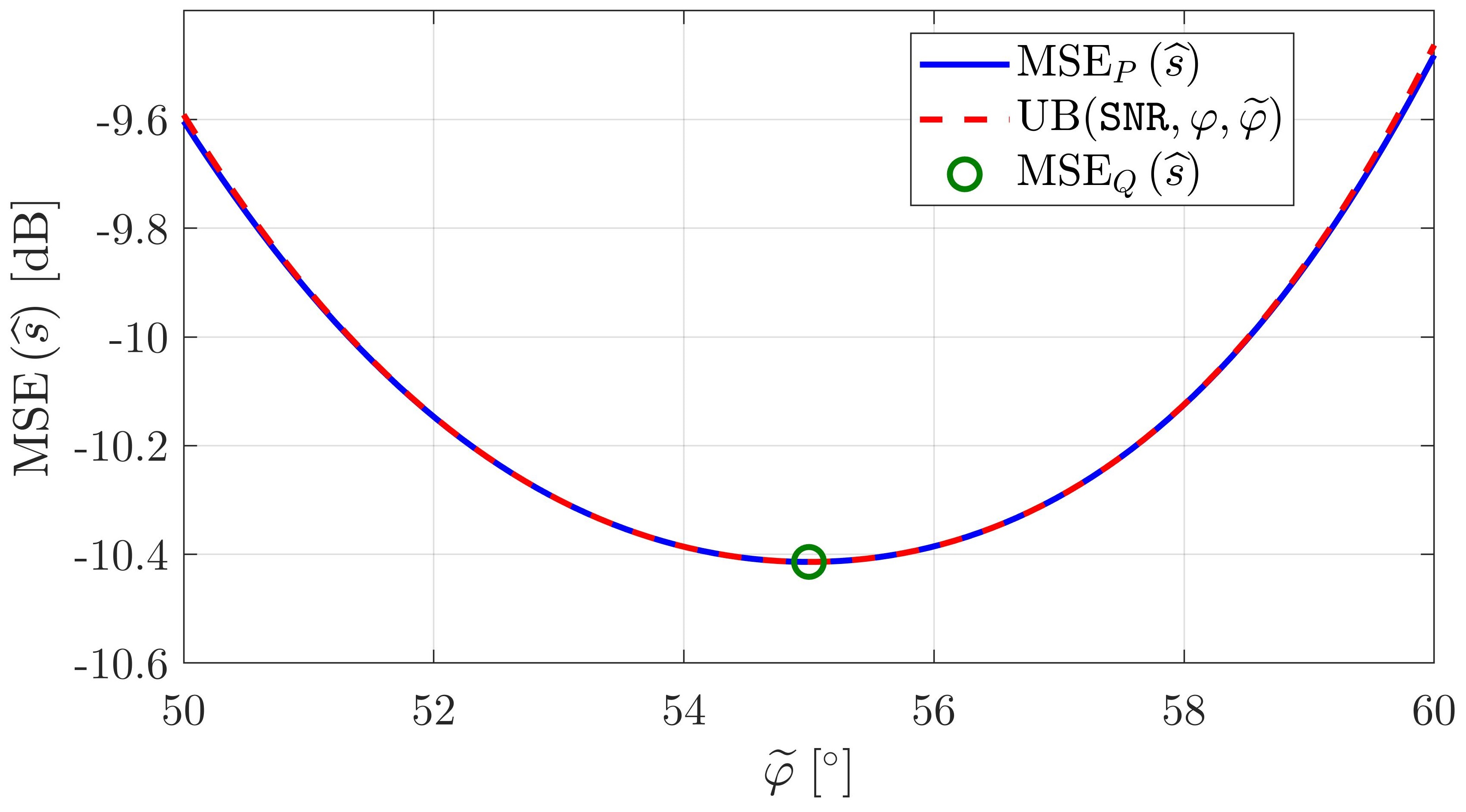}\vspace{-0.1cm}
\caption{MSE and the \delra{bilateral}\addra{upper} bound\addra{ \eqref{Gaussianexamplebound}} vs.~$\delra{\gamma}\addra{\widetilde{\varphi}}$ in ``\delra{variance}\addra{DOA}-mismatched'' \delra{Gaussian mean} estimation for $\delra{\sigma_P^2=1}\addra{\varphi=55^{\circ}, \snr=10}$\delra{ and $N=10$}.\addra{ Here, $\MSE_Q\left(\widehat{\rnds}\right)$ is independent of $(\varphi,\widetilde{\varphi})$.}}
\label{fig:Gaussian_toy_example2}\vspace{-0.55cm}
\end{figure}
Moreover, in this case the CSD is given by
\begin{equation*}
    \chi^2\left(P||Q\right)=\frac{\gamma^2}{\sqrt{2\gamma^2-1}}-1\triangleq c^2(\delra{\gamma}\addra{\snr,\varphi,\widetilde{\varphi}}), \delra{\quad}\addra{\;\;} \forall\gamma>\frac{1}{\sqrt{2}},
\end{equation*}
where $\gamma\delra{\triangleq\frac{\sigma_Q}{\sigma_P}}$\delra{, and is independent of $N$.}\addra{ is shorthand for $\gamma(\snr,\varphi,\widetilde{\varphi})$. Thus, using \eqref{bilateralbound2},}\delra{ We therefore obtain}
\begin{equation}\label{Gaussianexamplebound}
    \delra{\frac{\sigma_Q^2(1 - c(\gamma)\sqrt{2})}{N} \leq} \MSE_P\left(\widehat{\delra{\theta}\addra{\rnds}}\right) \leq \delra{\frac{\sigma_Q^2(1 + c(\gamma)\sqrt{2})}{N}}\addra{\frac{(1 + c(\snr,\varphi,\widetilde{\varphi})\sqrt{2})}{1+\snr}\triangleq {\rm UB}(\snr,\varphi,\widetilde{\varphi}),}\delra{.}
\end{equation}
\addra{where an obvious lower bound is $\MSE_Q\left(\widehat{\rnds}\right)$, since the MSE is minimized for the true DOA. Fig.~\ref{fig:Gaussian_toy_example2} shows the MSE under $P$ and the upper bound \eqref{Gaussianexamplebound} (i.e., \eqref{bilateralbound2}) vs.\ the mismatched DOA $\widetilde{\varphi}$ for an angular uncertainty of $10^{\circ}$ centered at the true DOA $\varphi=55^{\circ}$ for $\snr=10$ dB; the tightness of \eqref{bilateralbound2} is evident.

An intuitive interpretation of the above is that $Q_{\data}$ represents an ``overly optimistic'' point of view ($\gamma\leq1$), where the assumed DOA is exact: $\widetilde{\varphi}=\varphi$. Thus, the upper bound is closer to the true performance curve, and the lower bound becomes less informative. Clearly, here $\MSE_Q\left(\widehat{\addra{\rnds}}\right)\leq \MSE_P\left(\widehat{\addra{\rnds}}\right)$ for any (DOA) mismatch. Furthermore, $\MSE_Q\left(\widehat{\addra{\rnds}}\right)$ is greater than the lower bound \eqref{bilateralbound0}. Evidently, this example reveals that the bounds \eqref{bilateralbound0} and \eqref{bilateralbound2} are not always \emph{simultaneously} informative. However, perhaps surprisingly, in some situations they are, as demonstrated in the next section.
}

\section{\addra{Representative }Application\delra{s} of the Bound}\label{sec:applications}
\addra{We now consider a representative application of the bound in a nonBayesian framework (in contrast to that of Section \ref{subsec:examplecalculation}), which showcases bilateral tightness, as well as an improvement over the MCRB in the non-asymptotic regime.}

\addra{C}onsider\addra{ the} time-of-arrival (TOA) estimation\addra{ problem}, which is instrumental in a host of engineering applications, e.g., \cite{taylor2001ultra,falsi2006time,jansen2010terahertz}. Specifically, we focus on the case study of a mismatched waveform considered in \cite{roemer2020misspecified}, which is a special case of \eqref{signal_in_noise}, with $\theta=\tau$ (i.e., $K=1$), $\rvecv\sim\setN(\rveczero,\sigma^2\matI\addra{_M})$, and
\begin{equation*}
h_m(\tau)=\left.e^{-\left(\frac{t-\tau}{T_p}\right)^2}\right|_{t=mT_s}\triangleq h_m(\tau,T_p),\; m\in\{1,\ldots,M\},    
\end{equation*}
where $\tau$ is the unknown TOA, $h_m(\tau,T_p)$ is the $m$-th sample of a $\tau$-shifted Gaussian pulse with pulse width $T_p$, and $T_s$ is the sampling period, hence $MT_s$ is the observation interval.

While the true model of the observation $\rvecx$ is $P_{\data}=\setN(\rvech(\tau,T_p),\sigma^2\matI\addra{_M})$, if there is imprecise knowledge of the pulse width,\footnote{Such a mismatch can occur, for example, in ultrasound \cite{jensen1994nonparametric}.} and it is assumed to be $T_Q$, we have $Q_{\data}=\setN(\rvech(\tau,T_Q),\sigma^2\matI\addra{_M})$. When the\delra{ TOA} estimator\addra{ $\widehat{\tau}$} is designed based on $T_Q$, and the system designer is aware of the potential mismatch due to some inherent physical uncertainty, performance guarantees (upper bounds) and fundamental limitation (lower bounds) of the actual MSE can be of high practical value.\addra{

\begin{remark}
\addra{Although we do not use the looser version of the bound (Corollary \ref{corollary1}), we now show that in this problem, the CSD of the data distributions can be computed in closed-form. }It is known that for two multivariate normal distributions $P=\setN(\vecmu_P,\matLambda), Q=\setN(\vecmu_Q,\matLambda)$ \cite{ali1966general},
\begin{equation}
    D_f(P||Q)=D_f(\setN(0,1)||\setN(\delta_{\matLambda}(\vecmu_P,\vecmu_Q),1)),
\end{equation}
where\addraisit{ $\delta^2_{\matLambda}(\vecmu_P,\vecmu_Q)\triangleq\tp{(\vecmu_P-\vecmu_Q)}\matLambda^{-1}(\vecmu_P-\vecmu_Q)$
is the Mahalanobis generalized distance.} Hence, in our TOA estimation problem, we \delra{immediately }have  (e.g., [26])
\begin{equation}\label{chisqrdivTOA}
    \chi^2\left(P_{\data}||Q_{\data}\right)=\exp\left\{\tfrac{\|\rvech(\tau,T_P)-\rvech(\tau,T_Q)\|_2^2}{\sigma^{2}}\right\}-1.
\end{equation}
However, as explained in [21], for Gaussian pulses, the squared-norm in \eqref{chisqrdivTOA} can be approximated (with exponentially vanishing approximation errors) by an integral over the whole real line, after which a trivial change of integration variable shows that the squared norm in \eqref{chisqrdivTOA} is independent of $\tau$. 
\end{remark}

It is well-known that in this problem, at low signal-to-noise ratios (SNRs), the estimation error (of any reasonable estimator) is uniformly distributed on the uncertainty time-interval. Furthermore, at high SNRs, a cross-correlation-based estimator (CCE) is normally distributed around the true TOA, where the optimal CCE attains the minimal attainable variance. Thus, we expect that the CSD between the \emph{error} distributions of two slightly different CCEs will not differ significantly.
}
\begin{figure}[t!]
\centering
\includegraphics[width=0.875\columnwidth]{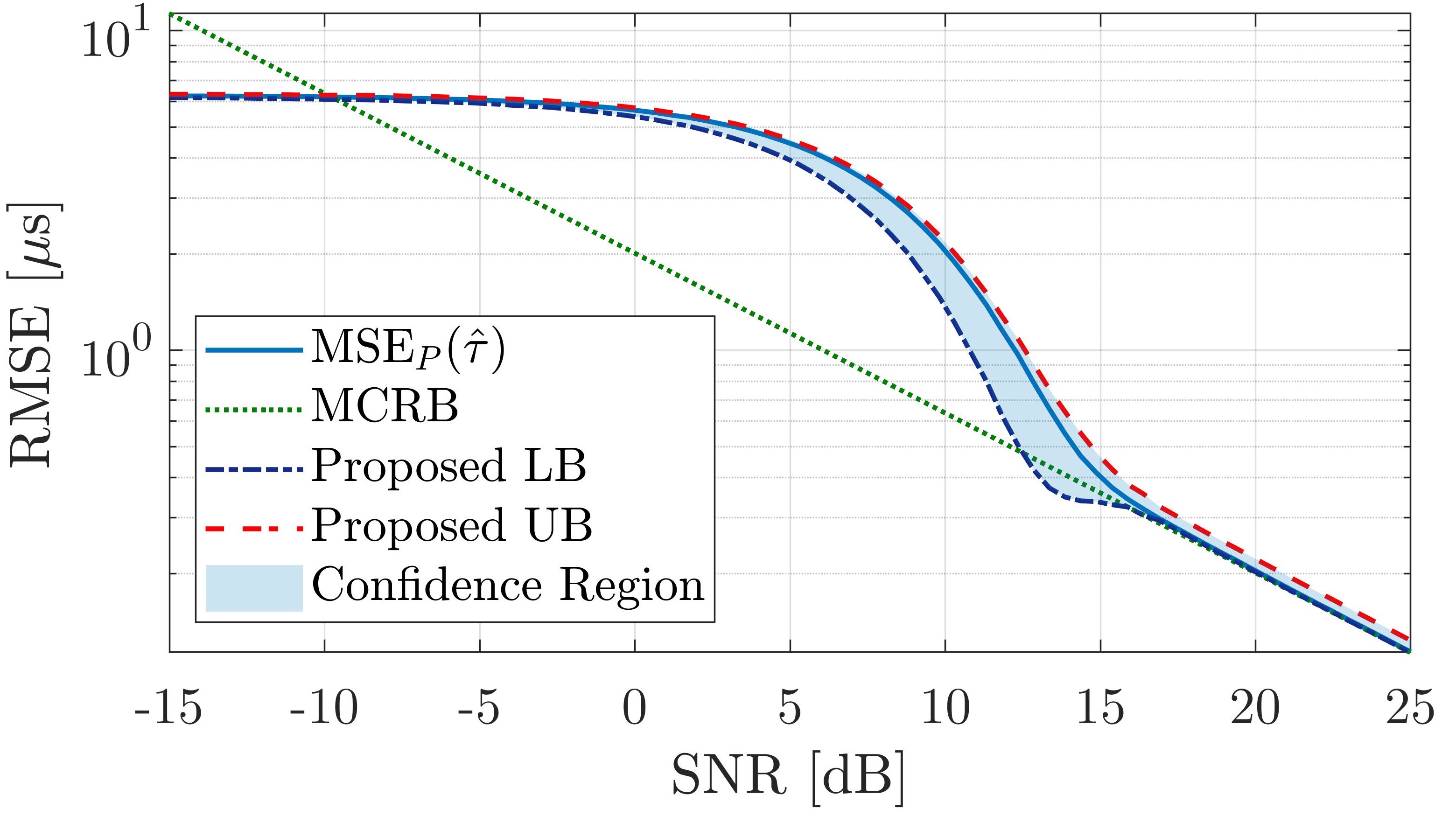}\vspace{-0.2cm}
\caption{Root MSE (RMSE) vs.\ SNR in TOA estimation, for $T_s=10^{-2}\mu{\rm{s}}, T_P=2\mu{\rm{s}}, T_Q=1.1\cdot T_P, M=2\cdot10^3$ and $\tau\sim{\rm{Unif}(-5,5)\mu{\rm{s}}}$.}
\label{fig:TOA_estimation}\vspace{-0.6cm}
\end{figure}

To compute the CSD between the distributions of the squared errors resulting from $P_{\data}$ and $Q_{\data}$ (Theorem \ref{theorem1}), we use the data-dependent partition divergence estimator \cite[Eq. 6]{wang2005divergence}. For the lower bound, we also use the (trivial) fact that the MSE is a monotonic nonincreasing function of the SNR, namely that ${\rm{MSE}}(\snr_1)\leq {\rm{MSE}}(\snr_2)$ for all $\snr_1\geq \snr_2$. Hence, a refinement of the lower bound \eqref{bilateralbound0} in this case is
\begin{equation}\label{refinedlowerbound}
{\rm{LB}}(\snr)\triangleq\max\{ {\rm{LB}}_{{\rm Thm 1}}(\varrho^2): \varrho^2\geq\snr \},
\end{equation}
where ${\rm{LB}}_{{\rm Thm 1}}(\varrho^2)$ denotes \eqref{refinedlowerbound} at an SNR level of $\varrho^2$.

Fig.~\ref{fig:TOA_estimation} presents the MSE vs.\ the SNR of the CCE, designed based on $T_Q=1.1\cdot T_P$ (namely, erroneously assuming $\rvecx\sim Q_{\data}$), the MCRB \cite[Eq.~14]{roemer2020misspecified}, \addra{and }the proposed\addra{ (refined) lower \eqref{refinedlowerbound} and} upper\addra{ \eqref{bilateralbound2}} bound\addra{s} (\addra{LB and }UB\addra{, respectively})\addra{.}\delra{ \eqref{upperboundweak}, and the improved lower bound (LB), defined as the maximum between \eqref{lowerboundweak} and the MCRB.} Evidently, the proposed LB is tighter than the MCRB at the low SNR regime, \addra{where the MCRB is not only uninformative, but no longer serves as a legitimate lower bound. In the transition region, the LB and UB}\delra{and} satisfactorily capture\delra{s} the threshold phenomenon (e.g., \cite{weiss1983fundamental,weinstein1984fundamental}). Moreover, the UB\addra{ \eqref{bilateralbound2}} provides the guaranteed accuracy despite the mismatch\delra{ in a wide range of SNRs, and diverges (as expected) at high SNR, where the MCRB is tight}.

\delra{Although upper bounds on the MSE have been considered for some (mismatch-free) special cases, e.g., \cite{seidman1968upper,timor1970upper,zakai1972lower,hawkes1976upper,ephraim1992lower,schniter2000bounds,belliardo2020achieving}, we are currently not aware of an upper bound on the MSE, for a given estimator, under model mismatch. Thus, improving upon the proposed upper bound in the high SNR regime is an attractive topic for future research.}

 \vspace{-0.1cm}
\section{Concluding Remarks}\label{sec:conclusion}
We develop a bilateral bound on MSE that is applicable to a general (not necessarily unbiased) estimator derived under a mismatched model. The bound provides performance guarantees for the operation of estimators that are designed to operate in one setting, but are then applied in a different one.

An interesting direction for future research is the potential applications of the bound in the context of machine learning, such as supervised regression, where the presence and assumption of model mismatch is ubiquitous.

\newpage
\bibliographystyle{IEEEtran}
\bibliography{./refs}

\end{document}

%% file: glossary.tex
\newglossaryentry{simo}
{
  name={SIMO},
  description={single-input multiple-output},
  first={\glsentrydesc{simo} (\glsentrytext{simo})}
}

\newglossaryentry{mmse}
{
  name={MMSE},
  description={minimum mean-square error},
  first={\glsentrydesc{mmse} (\glsentrytext{mmse})}
}

\newglossaryentry{lmmse}
{
  name={LMMSE},
  description={linear minimum mean-square error},
  first={\glsentrydesc{lmmse} (\glsentrytext{lmmse})}
}

\newglossaryentry{mf}
{
  name={MF},
  description={matched filtering},
  first={\glsentrydesc{mf} (\glsentrytext{mf})}
}

\newglossaryentry{qlmmse}
{
  name={QLMMSE},
  description={quasilinear MMSE},
  first={\glsentrydesc{qlmmse} (\glsentrytext{qlmmse})}
}

\newglossaryentry{na}
{
  name={NA},
  description={normal approximation},
  first={\glsentrydesc{na} (\glsentrytext{na})},
  plural={NAs},
  descriptionplural={normal approximations},
  firstplural={\glsentrydescplural{na} (\glsentryplural{na})}
}

\newglossaryentry{zf}
{
  name={ZF},
  description={zero-forcing},
  first={\glsentrydesc{zf} (\glsentrytext{zf})}
}

\newglossaryentry{rzf}
{
  name={RZF},
  description={regularized zero-forcing},
  first={\glsentrydesc{rzf} (\glsentrytext{rzf})}
}

\newglossaryentry{sir}
{
  name={SIR},
  description={signal-to-interference ratio},
  first={\glsentrydesc{sir} (\glsentrytext{sir})}
}

\newglossaryentry{snr}
{
  name={SNR},
  description={signal-to-noise ratio},
  first={\glsentrydesc{snr} (\glsentrytext{snr})}
}

\newglossaryentry{mse}
{
  name={MSE},
  description={mean-square error},
  first={\glsentrydesc{mse} (\glsentrytext{mse})}
}
\newglossaryentry{mmmse}
{
  name={M-MMSE},
  description={multicell MMSE},
  first={\glsentrydesc{mmmse} (\glsentrytext{mmmse})}
}

\newglossaryentry{ls}
{
  name={LS},
  description={least-squares},
  first={\glsentrydesc{ls} (\glsentrytext{ls})}
}

\newglossaryentry{mr}
{
  name={MR},
  description={maximum ratio},
  first={\glsentrydesc{mr} (\glsentrytext{mr})}
}

\newglossaryentry{ue}
{
  name={UE},
  description={user equipment},
  first={\glsentrydesc{ue} (\glsentrytext{ue})},
  plural={UEs},
  descriptionplural={user equipments},
  firstplural={\glsentrydescplural{ue} (\glsentryplural{ue})}
}

\newglossaryentry{ap}
{
  name={AP},
  description={access point},
  first={\glsentrydesc{ap} (\glsentrytext{ap})},
  plural={APs},
  descriptionplural={access points},
  firstplural={\glsentrydescplural{ap} (\glsentryplural{ap})}
}

\newglossaryentry{cpu}
{
  name={CPU},
  description={central processing unit},
  first={\glsentrydesc{cpu} (\glsentrytext{cpu})},
  plural={CPUs},
  descriptionplural={central processing units},
  firstplural={\glsentrydescplural{cpu} (\glsentryplural{cpu})}
}

\newglossaryentry{bs}
{
  name={BS},
  description={base station},
  first={\glsentrydesc{bs} (\glsentrytext{bs})},
  plural={BSs},
  descriptionplural={base stations},
  firstplural={\glsentrydescplural{bs} (\glsentryplural{bs})}
}

\newglossaryentry{ostbc}
{
  name={OSTBC},
  description={orthogonal space-time block code},
  first={\glsentrydesc{ostbc} (\glsentrytext{ostbc})},
  plural={OSTBCs},
  descriptionplural={orthogonal space-time block codes},
  firstplural={\glsentrydescplural{ostbc} (\glsentryplural{ostbc})}
}

\newglossaryentry{biawgn}
{
  name={bi-AWGN},
  description={binary-input additive white Gaussian noise},
  first={\glsentrydesc{biawgn} (\glsentrytext{biawgn})}
}

\newglossaryentry{flnf}
{
  name={FLNF},
  description={fixed-length no-feedback},
  first={\glsentrydesc{flnf} (\glsentrytext{flnf})}
}

\newglossaryentry{crc}
{
  name={CRC},
  description={cyclic redundancy check},
  first={\glsentrydesc{crc} (\glsentrytext{crc})}
}

\newglossaryentry{bsc}
{
  name={BSC},
  description={binary symmetric channel},
  first={\glsentrydesc{bsc} (\glsentrytext{bsc})}
  plural={BSCs},
  descriptionplural={binary symmetric channels},
  firstplural={\glsentrydescplural{bsc} (\glsentryplural{bsc})}
}

\newglossaryentry{bec}
{
  name={BEC},
  description={binary-erasure channel},
  first={\glsentrydesc{bec} (\glsentrytext{bec})}
}
\newglossaryentry{awgn}
{
  name={AWGN},
  description={additive white Gaussian noise},
  first={\glsentrydesc{awgn} (\glsentrytext{awgn})}
}
\newglossaryentry{3gpp}
{
  name={3GPP},
  description={3rd generation partnership project},
  first={\glsentrydesc{3gpp} (\glsentrytext{3gpp})}
}

\newglossaryentry{dl}
{
  name={DL},
  description={downlink},
  first={\glsentrydesc{dl} (\glsentrytext{dl})}
}

\newglossaryentry{LED}
{
  name={LED},
  description={light emitting diode},
  first={\glsentrydesc{LED} (\glsentrytext{LED})},
  plural={LEDs},
  descriptionplural={light emitting diodes},
  firstplural={\glsentrydescplural{LED} (\glsentryplural{LED})}
}

\newglossaryentry{lte}
{
  name={LTE},
  description={long term evolution},
  first={\glsentrydesc{lte} (\glsentrytext{lte})}
}
\newglossaryentry{ofdm}
{
  name={OFDM},
  description={orthogonal frequency-division multiplexing},
  first={\glsentrydesc{ofdm} (\glsentrytext{ofdm})}
}

\newglossaryentry{ul}
{
  name={UL},
  description={uplink},
  first={\glsentrydesc{ul} (\glsentrytext{ul})}
}
\newglossaryentry{rb}
{
  name={RB},
  description={resource block},
  first={\glsentrydesc{rb} (\glsentrytext{rb})},
  plural={RBs},
  descriptionplural={resource blocks},
  firstplural={\glsentrydescplural{rb} (\glsentryplural{rb})}
}
\newglossaryentry{cu}
{
  name={CU},
  description={channel use},
  first={\glsentrydesc{cu} (\glsentrytext{cu})},
  plural={CUs},
  descriptionplural={channel uses},
  firstplural={\glsentrydescplural{cu} (\glsentryplural{cu})}
}
\newglossaryentry{tti}
{
  name={TTI},
  description={transmission time interval},
  first={\glsentrydesc{tti} (\glsentrytext{tti})},
  plural={TTI's},
  descriptionplural={transmission time intervals},
  firstplural={\glsentrydescplural{tti} (\glsentryplural{tti})}
}
\newglossaryentry{iid}
{
  name={i.i.d.},
  description={independent and identically distributed},
  first={\glsentrydesc{iid} (\glsentrytext{iid})}
}
\newglossaryentry{psd}
{
  name={PSD},
  description={power spectral density},
  first={\glsentrydesc{psd} (\glsentrytext{psd})}
}

\newglossaryentry{mimo}
{
  name={MIMO},
  description={multiple-input multiple-output},
  first={\glsentrydesc{mimo} (\glsentrytext{mimo})}
}

\newglossaryentry{bler}
{
  name={BLER},
  description={block error probability},
  first={\glsentrydesc{bler} (\glsentrytext{bler})}
}
\newglossaryentry{mtc}
{
  name={MTC},
  description={machine-type communication},
  first={\glsentrydesc{mtc} (\glsentrytext{mtc})}
}
\newacronym{csi}{CSI}{channel state information}
\newglossaryentry{dvbt}
{
  name={DVB-T},
  description={terrestrial digital video broadcasting},
  first={\glsentrydesc{dvbt} (\glsentrytext{dvbt})}
}
\newglossaryentry{pat}
{
  name={PAT},
  description={pilot-assisted transmission},
  first={\glsentrydesc{pat} (\glsentrytext{pat})}
}
\newglossaryentry{ml}
{
  name={ML},
  description={maximum likelihood},
  first={\glsentrydesc{ml} (\glsentrytext{ml})}
}
\newglossaryentry{dt}
{
  name={DT},
  description={dependence-testing},
  first={\glsentrydesc{dt} (\glsentrytext{dt})}
}
\newglossaryentry{ustm}
{
  name={USTM},
  description={unitary space-time modulation},
  first={\glsentrydesc{ustm} (\glsentrytext{ustm})}
}
\newglossaryentry{siso}
{
  name={SISO},
  description={single-input single-output},
  first={\glsentrydesc{siso} (\glsentrytext{siso})}
}
\newglossaryentry{snn}
{
  name={SNN},
  description={scaled nearest-neighbor},
  first={\glsentrydesc{snn} (\glsentrytext{snn})}
}

\newglossaryentry{snnd}
{
  name={SNND},
  description={scaled nearest-neighbor decoding},
  first={\glsentrydesc{snnd} (\glsentrytext{snnd})}
}
\newglossaryentry{rcus}
{
  name={RCUs},
  description={random-coding union bound with parameter $s$},
  first={\glsentrydesc{rcus} (\glsentrytext{rcus})}
}
\newglossaryentry{osd}
{
  name={OSD},
  description={ordered statistics decoding},
  first={\glsentrydesc{osd} (\glsentrytext{osd})}
}
\newglossaryentry{llr}
{
  name={LLR},
  description={log-likelihood ratio},
  first={\glsentrydesc{llr} (\glsentrytext{llr})}
   plural={LLR's},
  descriptionplural={log-likelihood ratios},
  firstplural={\glsentrydescplural{llr} (\glsentryplural{llr})}
}
\newglossaryentry{qpsk}
{
  name={QPSK},
  description={quaternary phase shift keying},
  first={\glsentrydesc{qpsk} (\glsentrytext{qpsk})}
}
\newglossaryentry{nlos}
{
  name={NLOS},
  description={non-line-of-sight},
  first={\glsentrydesc{nlos} (\glsentrytext{nlos})}
}
\newglossaryentry{los}
{
  name={LOS},
  description={line-of-sight},
  first={\glsentrydesc{los} (\glsentrytext{los})}
}
\newglossaryentry{mgf}
{
  name={MGF},
  description={moment-generating function},
  first={\glsentrydesc{mgf} (\glsentrytext{mgf})}
}
\newglossaryentry{cgf}
{
  name={CGF},
  description={cumulant-generating function},
  first={\glsentrydesc{cgf} (\glsentrytext{cgf})},
  plural={CGFs},
  descriptionplural={cumulant-generating functions},
  firstplural={\glsentrydescplural{cgf} (\glsentryplural{cgf})}
}
\newglossaryentry{pdf}
{
  name={PDF},
  description={probability density function},
  first={\glsentrydesc{pdf} (\glsentrytext{pdf})},
   plural={PDF's},
  descriptionplural={probability density functions},
  firstplural={\glsentrydescplural{pdf} (\glsentryplural{pdf})}
}

\newglossaryentry{ccdf}
{
  name={CCDF},
  description={complementary cumulative distribution function},
  first={\glsentrydesc{ccdf} (\glsentrytext{ccdf})}
}

\newglossaryentry{cdf}
{
  name={CDF},
  description={cumulative distribution function},
  first={\glsentrydesc{cdf} (\glsentrytext{cdf})}
}

\newglossaryentry{gmi}
{
  name={GMI},
  description={generalized mutual information},
  first={\glsentrydesc{gmi} (\glsentrytext{gmi})}
}

\newglossaryentry{arq}
{
  name={ARQ},
  description={automatic repeat request},
  first={\glsentrydesc{arq} (\glsentrytext{arq})}
}

\newglossaryentry{harq}
{
  name={HARQ},
  description={hybrid automatic repeat request},
  first={\glsentrydesc{harq} (\glsentrytext{harq})}
}

\newglossaryentry{fbl}
{
  name={FBL-NF},
  description={fixed blocklength with no feedback},
  first={\glsentrydesc{fbl} (\glsentrytext{fbl})}
}
\newglossaryentry{ssnf}
{
  name={SS-NF},
  description={single-shot no-feedback},
  first={\glsentrydesc{ssnf} (\glsentrytext{ssnf})}
}

\newacronym{urllc}{URLLC}{ultra-reliable low-latency communications}

\newglossaryentry{vlsf}
{
  name={VLSF},
  description={Variable-length stop-feedback},
  first={\glsentrydesc{vlsf} (\glsentrytext{vlsf})}
}

\newglossaryentry{vlnsf}
{
  name={VLNSF},
  description={variable-length noisy stop feedback},
  first={\glsentrydesc{vlnsf} (\glsentrytext{vlnsf})}
}

\newglossaryentry{dmt}
{
  name={DMT},
  description={diversity-multiplexing tradeoff},
  first={\glsentrydesc{dmt} (\glsentrytext{dmt})}
}

\newglossaryentry{dmdt}
{
  name={DMDT},
  description={diversity-multiplexing-delay tradeoff},
  first={\glsentrydesc{dmdt} (\glsentrytext{dmdt})}
}

\newglossaryentry{tddt}
{
  name={TDDT},
  description={throughput-diversity-delay tradeoff},
  first={\glsentrydesc{tddt} (\glsentrytext{tddt})}
}

\newglossaryentry{tdd}
{
  name={TDD},
  description={time-division duplex},
  first={\glsentrydesc{tdd} (\glsentrytext{tdd})}
}

\newglossaryentry{stbc}
{
  name={STBC},
  description={space-time block-codes},
  first={\glsentrydesc{stbc} (\glsentrytext{stbc})},
  plural={STBC's},
  descriptionplural={space-time block-codes},
  firstplural={\glsentrydescplural{stbc} (\glsentryplural{stbc})}
}

\newglossaryentry{harqir}
{
  name={HARQ-IR},
  description={hybrid automatic repeat request with incremental redundancy},
  first={\glsentrydesc{harqir} (\glsentrytext{harqir})}
}

\newglossaryentry{harqcc}
{
  name={HARQ-CC},
  description={hybrid automatic repeat request with chase combining},
  first={\glsentrydesc{harqcc} (\glsentrytext{harqcc})}
}

\newglossaryentry{wp1}
{
  name={w.p.1.},
  description={with probability one},
  first={\glsentrydesc{wp1} (\glsentrytext{wp1})}
}

\newglossaryentry{vlff}
{
  name={VLFF},
  description={variable-length full feedback},
  first={\glsentrydesc{vlff} (\glsentrytext{vlff})}
}

\newglossaryentry{dmc}
{
  name={DMC},
  description={discrete memoryless channel},
  first={\glsentrydesc{dmc} (\glsentrytext{dmc})}
  plural={DMCs},
  descriptionplural={discrete memoryless channels},
  firstplural={\glsentrydescplural{dmc} (\glsentryplural{dmc})}
}

\newglossaryentry{dnn}
{
  name={DNN},
  description={deep neural network},
  first={\glsentrydesc{dnn} (\glsentrytext{dnn})}
  plural={DNNs},
  descriptionplural={deep neural networks},
  firstplural={\glsentrydescplural{dnn} (\glsentryplural{dnn})}
}

\newglossaryentry{nn}
{
  name={NN},
  description={neural network},
  first={\glsentrydesc{nn} (\glsentrytext{nn})}
  plural={NNs},
  descriptionplural={neural networks},
  firstplural={\glsentrydescplural{nn} (\glsentryplural{nn})}
}

\newglossaryentry{cnn}
{
  name={CNN},
  description={convolutional neural network},
  first={\glsentrydesc{cnn} (\glsentrytext{cnn})}
  plural={CNNs},
  descriptionplural={convolutional neural networks},
  firstplural={\glsentrydescplural{cnn} (\glsentryplural{cnn})}
}

\newglossaryentry{ber}
{
  name={BER},
  description={bit error rate},
  first={\glsentrydesc{ber} (\glsentrytext{ber})}
}

\newglossaryentry{scss}
{
  name={SCSS},
  description={\emph{single-channel} source separation},
  first={\glsentrydesc{scss} (\glsentrytext{scss})}
}

\newglossaryentry{fft}
{
  name={FFT},
  description={fast Fourier transform},
  first={\glsentrydesc{fft} (\glsentrytext{fft})}
}

\newglossaryentry{soi}
{
  name={SOI},
  description={signal of interest},
  first={\glsentrydesc{soi} (\glsentrytext{soi})}
}

\newglossaryentry{map}
{
  name={MAP},
  description={maximum \emph{a posteriori}},
  first={\glsentrydesc{map} (\glsentrytext{map})}
}

\newglossaryentry{usc}
{
  name={TDC},
  description={``temporal-diversity" condition},
  first={\glsentrydesc{usc} (\glsentrytext{usc})}
}

\newglossaryentry{qam}
{
  name={QAM},
  description={quadrature amplitude modulation},
  first={\glsentrydesc{qam} (\glsentrytext{qam})}
}

\newglossaryentry{mit}
{
  name={MIT},
  description={Massachusetts Institute of Technology},
  first={\glsentrydesc{mit} (\glsentrytext{mit})}
}

\newglossaryentry{rf}
{
  name={RF},
  description={radio-frequency},
  first={\glsentrydesc{rf} (\glsentrytext{rf})}
}

%% file: isit2023_arxiv_version.bbl
\begin{thebibliography}{10}
\providecommand{\url}[1]{#1}
\csname url@samestyle\endcsname
\providecommand{\newblock}{\relax}
\providecommand{\bibinfo}[2]{#2}
\providecommand{\BIBentrySTDinterwordspacing}{\spaceskip=0pt\relax}
\providecommand{\BIBentryALTinterwordstretchfactor}{4}
\providecommand{\BIBentryALTinterwordspacing}{\spaceskip=\fontdimen2\font plus
\BIBentryALTinterwordstretchfactor\fontdimen3\font minus
  \fontdimen4\font\relax}
\providecommand{\BIBforeignlanguage}[2]{{%
\expandafter\ifx\csname l@#1\endcsname\relax
\typeout{** WARNING: IEEEtran.bst: No hyphenation pattern has been}%
\typeout{** loaded for the language `#1'. Using the pattern for}%
\typeout{** the default language instead.}%
\else
\language=\csname l@#1\endcsname
\fi
#2}}
\providecommand{\BIBdecl}{\relax}
\BIBdecl

\bibitem{kay1993fundamentals}
S.~M. Kay, \emph{Fundamentals of statistical signal processing: estimation
  theory}.\hskip 1em plus 0.5em minus 0.4em\relax Prentice-Hall, Inc., 1993.

\bibitem{van2004detection}
H.~L. Van~Trees, \emph{Detection, estimation, and modulation theory, part I:
  detection, estimation, and linear modulation theory}.\hskip 1em plus 0.5em
  minus 0.4em\relax John Wiley \& Sons, 2004.

\bibitem{barankin1949locally}
E.~W. Barankin, ``Locally best unbiased estimates,'' \emph{The Annals of
  Mathematical Statistics}, vol.~20, no.~4, pp. 477--501, 1949.

\bibitem{ziv1969some}
J.~Ziv and M.~Zakai, ``Some lower bounds on signal parameter estimation,''
  \emph{IEEE Trans. Inf. Theory}, vol.~15, no.~3, pp. 386--391, 1969.

\bibitem{abel1993bound}
J.~S. Abel, ``A bound on mean-square-estimate error,'' \emph{IEEE Trans. Inf.
  Theory}, vol.~39, no.~5, pp. 1675--1680, 1993.

\bibitem{weiss1985lower}
A.~J. Weiss and E.~Weinstein, ``A lower bound on the mean-square error in
  random parameter estimation (corresp.),'' \emph{IEEE Trans. Inf. Theory},
  vol.~31, no.~5, pp. 680--682, 1985.

\bibitem{todros2010generalA}
K.~Todros and J.~Tabrikian, ``General classes of performance lower bounds for
  parameter estimation—{P}art {I}: Non-{B}ayesian bounds for unbiased
  estimators,'' \emph{IEEE Trans. Inf. Theory}, vol.~56, no.~10, pp.
  5045--5063, 2010.

\bibitem{todros2010generalB}
------, ``General classes of performance lower bounds for parameter
  estimation—{P}art {II}: {B}ayesian bounds,'' \emph{IEEE Trans. Inf.
  Theory}, vol.~56, no.~10, pp. 5064--5082, 2010.

\bibitem{fortunati2017performance}
S.~Fortunati, F.~Gini, M.~S. Greco, and C.~D. Richmond, ``Performance bounds
  for parameter estimation under misspecified models: Fundamental findings and
  applications,'' \emph{IEEE Signal Process. Mag.}, vol.~34, no.~6, pp.
  142--157, 2017.

\bibitem{huber1967under}
P.~J. Huber, ``The behavior of maximum likelihood estimates under nonstandard
  conditions,'' in \emph{Proceedings of the Fifth Berkeley Symposium on
  Mathematical Statistics and Probability: Weather Modification; University of
  California Press: Berkeley, CA, USA}, 1967, p. 221.

\bibitem{white1982maximum}
H.~White, ``Maximum likelihood estimation of misspecified models,''
  \emph{Econometrica: Journal of the econometric society}, pp. 1--25, 1982.

\bibitem{vuong1986cramer}
Q.~H. Vuong, ``{C}ram\'er-{R}ao bounds for misspecified models,'' \emph{Div. of
  the Humanities and Social Sci., California Inst. of Technol., Pasadena, CA,
  USA}, 1986.

\bibitem{xu2004bound}
W.~Xu, A.~B. Baggeroer, and K.~L. Bell, ``A bound on mean-square estimation
  error with background parameter mismatch,'' \emph{IEEE Trans. Inf. Theory},
  vol.~50, no.~4, pp. 621--632, 2004.

\bibitem{verdu2010mismatched}
S.~Verd{\'u}, ``Mismatched estimation and relative entropy,'' \emph{IEEE Trans.
  Inf. Theory}, vol.~56, no.~8, pp. 3712--3720, 2010.

\bibitem{fozunbal2010regret}
M.~Fozunbal, ``On regret of parametric mismatch in minimum mean square error
  estimation,'' in \emph{IEEE Int. Symp. Inf. Theory (ISIT)}, 2010, pp.
  1408--1412.

\bibitem{fritsche2015cramer}
C.~Fritsche, U.~Orguner, E.~{\"O}zkan, and F.~Gustafsson, ``On the
  {C}ram{\'e}r-{R}ao lower bound under model mismatch,'' in \emph{Proc. of
  ICASSP}, 2015, pp. 3986--3990.

\bibitem{fortunati2015lower}
S.~Fortunati, M.~S. Greco, and F.~Gini, ``A lower bound for the mismatched
  maximum likelihood estimator,'' in \emph{IEEE Radar Conference (RadarCon)},
  2015, pp. 0180--0185.

\bibitem{richmond2015parameter}
C.~D. Richmond and L.~L. Horowitz, ``Parameter bounds on estimation accuracy
  under model misspecification,'' \emph{IEEE Trans. Signal Process.}, vol.~63,
  no.~9, pp. 2263--2278, 2015.

\bibitem{diong2017generalized}
M.~L. Diong, E.~Chaumette, and F.~Vincent, ``Generalized {B}arankin-type lower
  bounds for misspecified models,'' in \emph{Proc. of ICASSP}, 2017, pp.
  4466--4470.

\bibitem{pajovic2018misspecified}
M.~Pajovic, ``Misspecified {B}ayesian cram{\'e}r-rao bound for sparse
  {B}ayesian,'' in \emph{2018 IEEE Statistical Signal Processing Workshop
  (SSP)}, 2018, pp. 263--267.

\bibitem{roemer2020misspecified}
F.~Roemer, ``Misspecified {C}ramer-{R}ao bound for delay estimation with a
  mismatched waveform: A case study,'' in \emph{Proc. of ICASSP}, 2020, pp.
  5994--5998.

\bibitem{abed2021misspecified}
L.~T. Thanh, K.~Abed-Meraim, and N.~L. Trung, ``Misspecified {C}ramer--{R}ao
  bounds for blind channel estimation under channel order misspecification,''
  \emph{IEEE Trans. Signal Process.}, vol.~69, pp. 5372--5385, 2021.

\bibitem{fortunati2016misspecified}
S.~Fortunati, F.~Gini, and M.~S. Greco, ``The misspecified {C}ram{\'e}r-{R}ao
  bound and its application to scatter matrix estimation in complex
  elliptically symmetric distributions,'' \emph{IEEE Trans. Signal Process.},
  vol.~64, no.~9, pp. 2387--2399, 2016.

\bibitem{seidman1968upper}
L.~P. Seidman, ``An upper bound on average estimation error in nonlinear
  systems,'' \emph{IEEE Trans. Inf. Theory}, vol.~14, no.~2, pp. 243--250,
  1968.

\bibitem{timor1970upper}
U.~Timor, ``An upper bound on the estimation error in the threshold region,''
  \emph{IEEE Trans. Inf. Theory}, vol.~16, no.~6, pp. 692--699, 1970.

\bibitem{zakai1972lower}
M.~Zakai and J.~Ziv, ``Lower and upper bounds on the optimal filtering error of
  certain diffusion processes,'' \emph{IEEE Trans. Inf. Theory}, vol.~18,
  no.~3, pp. 325--331, 1972.

\bibitem{hawkes1976upper}
R.~M. Hawkes and J.~B. Moore, ``An upper bound on the mean-square error for
  {B}ayesian parameter estimators (corresp.),'' \emph{IEEE Trans. Inf. Theory},
  vol.~22, no.~5, pp. 610--615, 1976.

\bibitem{ephraim1992lower}
Y.~Ephraim and N.~Merhav, ``Lower and upper bounds on the minimum mean-square
  error in composite source signal estimation,'' \emph{IEEE Trans. Inf.
  Theory}, vol.~38, no.~6, pp. 1709--1724, 1992.

\bibitem{schniter2000bounds}
P.~Schniter and C.~R. Johnson, Jr., ``Bounds for the {MSE} performance of
  constant modulus estimators,'' \emph{IEEE Trans. Inf. Theory}, vol.~46,
  no.~7, pp. 2544--2560, 2000.

\bibitem{belliardo2020achieving}
F.~Belliardo and V.~Giovannetti, ``Achieving {H}eisenberg scaling with
  maximally entangled states: An analytic upper bound for the attainable
  root-mean-square error,'' \emph{Physical Review A}, vol. 102, no.~4, p.
  042613, 2020.

\bibitem{csiszar1967information}
I.~Csisz{\'a}r, ``Information-type measures of difference of probability
  distributions and indirect observation,'' \emph{Studia Sci. Math. Hungar.},
  vol.~2, pp. 229--318, 1967.

\bibitem{wu2017lecture}
Y.~Wu, ``Lecture notes on information-theoretic methods for high-dimensional
  statistics,'' \emph{Lecture Notes for ECE598YW (UIUC)}, vol.~16, 2017.

\bibitem{nielsen2013chi}
F.~Nielsen and R.~Nock, ``On the chi square and higher-order chi distances for
  approximating f-divergences,'' \emph{IEEE Signal Process. Lett.}, vol.~21,
  no.~1, pp. 10--13, 2013.

\bibitem{hammersley1950estimating}
J.~M. Hammersley, ``On estimating restricted parameters,'' \emph{Journal of the
  Royal Statistical Society. Series B (Methodological)}, vol.~12, no.~2, pp.
  192--240, 1950.

\bibitem{chapman1951minimum}
D.~G. Chapman and H.~Robbins, ``Minimum variance estimation without regularity
  assumptions,'' \emph{The Annals of Mathematical Statistics}, pp. 581--586,
  1951.

\bibitem{ross2010first}
S.~Ross, \emph{A first course in probability}.\hskip 1em plus 0.5em minus
  0.4em\relax Pearson, 2010.

\bibitem{csiszar1972class}
I.~Csisz{\'a}r, ``A class of measures of informativity of observation
  channels,'' \emph{Periodica Mathematica Hungarica}, vol.~2, no. 1-4, pp.
  191--213, 1972.

\bibitem{yeredor2018high}
A.~Yeredor, A.~Weiss, and A.~J. Weiss, ``High-order analysis of the efficiency
  gap for maximum likelihood estimation in nonlinear {G}aussian models,''
  \emph{IEEE Trans. Signal Process.}, vol.~66, no.~18, pp. 4782--4795, 2018.

\bibitem{cover1999elements}
T.~M. Cover and J.~A. Thomas, \emph{Elements of information theory}.\hskip 1em
  plus 0.5em minus 0.4em\relax John Wiley \& Sons, 1999.

\bibitem{isserlis1918formula}
L.~Isserlis, ``On a formula for the product-moment coefficient of any order of
  a normal frequency distribution in any number of variables,''
  \emph{Biometrika}, vol.~12, no. 1/2, pp. 134--139, 1918.

\bibitem{lehmann2006theory}
E.~L. Lehmann and G.~Casella, \emph{Theory of point estimation}.\hskip 1em plus
  0.5em minus 0.4em\relax Springer Science \& Business Media, 2006.

\bibitem{glave1972new}
F.~E. Glave, ``A new look at the {B}arankin lower bound,'' \emph{IEEE Trans.
  Inf. Theory}, vol.~18, no.~3, pp. 349--356, 1972.

\bibitem{tse2005fundamentals}
D.~Tse and P.~Viswanath, \emph{Fundamentals of wireless communication}.\hskip
  1em plus 0.5em minus 0.4em\relax Cambridge university press, 2005.

\bibitem{taylor2001ultra}
J.~D. Taylor, \emph{Ultra-Wideband Radar Systems}.\hskip 1em plus 0.5em minus
  0.4em\relax CRC press, 2001.

\bibitem{falsi2006time}
C.~Falsi, D.~Dardari, L.~Mucchi, and M.~Z. Win, ``Time of arrival estimation
  for {UWB} localizers in realistic environments,'' \emph{EURASIP J. Appl.
  Signal Process.}, vol. 2006, pp. 1--13, 2006.

\bibitem{jansen2010terahertz}
C.~Jansen, S.~Wietzke, O.~Peters, M.~Scheller, N.~Vieweg, M.~Salhi,
  N.~Krumbholz, C.~J{\"o}rdens, T.~Hochrein, and M.~Koch, ``Terahertz imaging:
  applications and perspectives,'' \emph{Applied optics}, vol.~49, no.~19, pp.
  E48--E57, 2010.

\bibitem{jensen1994nonparametric}
J.~A. Jensen and S.~Leeman, ``Nonparametric estimation of ultrasound pulses,''
  \emph{IEEE Trans. Biomed. Eng.}, vol.~41, no.~10, pp. 929--936, 1994.

\bibitem{ali1966general}
S.~M. Ali and S.~D. Silvey, ``A general class of coefficients of divergence of
  one distribution from another,'' \emph{J. Roy. Statist. Soc., Ser. B},
  vol.~28, pp. 131--142, 1966.

\bibitem{wang2005divergence}
Q.~Wang, S.~R. Kulkarni, and S.~Verd{\'u}, ``Divergence estimation of
  continuous distributions based on data-dependent partitions,'' \emph{IEEE
  Trans. Inf. Theory}, vol.~51, no.~9, pp. 3064--3074, 2005.

\bibitem{weiss1983fundamental}
A.~J. Weiss and E.~Weinstein, ``Fundamental limitations in passive time delay
  estimation--{P}art {I}: {N}arrow-band systems,'' \emph{IEEE Trans. Acoust.,
  Speech Signal Process.}, vol.~31, no.~2, pp. 472--486, 1983.

\bibitem{weinstein1984fundamental}
E.~Weinstein and A.~J. Weiss, ``Fundamental limitations in passive time-delay
  estimation--{P}art {II}: {W}ide-band systems,'' \emph{IEEE Trans. Acoust.,
  Speech Signal Process.,}, vol.~32, no.~5, pp. 1064--1078, 1984.

\end{thebibliography}
